\documentclass[aps,twocolumn,nofootinbib]{revtex4}
\usepackage[babel]{csquotes}
\usepackage{graphicx}
\usepackage{amsmath,amssymb}
\usepackage[colorlinks,citecolor=blue,linkcolor=blue,urlcolor=blue]{hyperref}
\usepackage{mathrsfs}
\usepackage{enumerate}
\def\be{\begin{equation}}
\def\ee{\end{equation}}

\begin{document}

\title{Fast Radio Bursts and White Hole Signals}

\author{Aur\'elien Barrau${}^a$, Carlo Rovelli${}^{b}$ and Francesca Vidotto${}^{b,c}$\vspace{.5em} }
\affiliation{${}^a$\ Laboratoire de Physique Subatomique et de Cosmologie, Universit\'e Grenoble-Alpes, CNRS-IN2P3 53, avenue des Martyrs, F-38026 Grenoble, France \vspace{.3em} \\
${}^b$\ CPT, Aix-Marseille Universit\'e, Universit\'e de Toulon, CNRS, 
and the Samy Maroun Research Center for Time, Space and the Quantum, 
Case 907, F-13288 Marseille, France. \vspace{.3em} \\
${}^c$\ Radboud University, Institute for Mathematics, Astrophysics and Particle Physics,
Mailbox 79, P.O. Box 9010, 6500 GL Nijmegen, The Netherlands.}

\date{\small\today}

\begin{abstract} 
\noindent  We estimate the size of a primordial black hole exploding today via a white hole transition, and the power in the resulting explosion, using a simple model.  We point out that Fast Radio Bursts, strong signals with millisecond duration, probably extragalactic and having unknown source, have wavelength not far from the expected size of the exploding hole.  We also discuss the possible higher energy components of the signal. 
\end{abstract}

\maketitle

\noindent

\section{The model}

The fate of the vast amount of matter fallen into black holes is unknown. A possibility investigated by numerous authors is that quantum gravity generates pressure (or weakens gravity) halting the collapse and triggering a bounce causing the black hole to explode \cite{Frolov:1979tu,Stephens1994,Frolov:1981mz, Hayward1994, HAJICEK2001,Modesto:2004xx,Ashtekar:2005cj,Modesto2006a, Hayward2006, Booth2006, Modesto2008,Modesto2008a,Hossenfelder:2009fc, Kavic2008b,Modesto2011a, frolov:BHclosed, Bardeen2014,Torres2014,Rama2014} possibly at a size  much larger than Planckian \cite{Giddings1992a, Bambi2014, Rovelli2014}. Lifetimes of stellar or galactic holes are far too long for us to have a chance to detect the resulting explosion.  But primordial black holes formed in the very early universe, if they exist \cite{Carr1975,Carr:1997cn,Capela2013,Barnacka2012}, could be exploding today. For a black hole of initial mass $m$, the hypothesis that the phenomenon prevents the firewall problem \cite{Almheiri:2012rt} implies a maximal lifetime shorter than the Hawking evaporation time \cite{Hawking}, but still of order $m^3$ in Planck units ($c=\hbar=G=1$). In \cite{Rovelli2014}, the signal emitted by a primordial black hole exploding today was estimated, under this maximal lifetime hypothesis,    to be in the Gev range. The phenomenology of such an event has been studied in \cite{Barrau2014}. For related suggestions see \cite{Bambi2013b, Gambini2013, Gambini2014, Gambini2014a, Barcelo2014, Barcelo2014a, Mersini-Houghton2014}. 

Later theoretical work on the gravitational field of such bouncing ``Planck star" has pointed out that quantum gravity effects might become relevant earlier, allowing for shorter blackhole lifetime  \cite{Haggard2014}. Classical general relativity outside the region of the hole is compatible with a black-to-white quantum transition. The black and white hole solutions of the Einstein equations can be glued and their singularities replaced by a \emph{finite} (in space and in time) non-classical tunnelling region.  An estimate of the time needed to exit the semiclassical regime yields a black hole lifetime of the order
\be
  \tau=4k\  m^2
  \label{tau}
\ee
in Planck units, where $k$ is estimated to $k=.05$ in \cite{Haggard2014}. Primordial black holes of initial mass around
\be
m = \sqrt{\frac{t_H}{4k}}\sim 1.2\times 10^{23}~{\rm kg}
\label{mass}
\ee 
where $t_H$ is the Hubble time, can therefore be expected to explode today.  The possibility of observing signals from  white holes was first pointed out long ago by Narlikar, Appa Rao and Dadhich in \cite{Narlikar1974}. 

A ``bounce" can take a cosmological time because of the general-relativistic time dilation: the proper time of an observer \emph{outside} the hole is cosmological, but the proper time of an observer bouncing with the star \emph{inside} the hole is very small (order $m$, namely the time light takes to cross the collapsing object).   

If this happens, most of the energy of the black hole is still present at explosion time, because Hawking radiation does not have the time to consume it. The exploding object should have a total energy of the order 
\be
E= mc^2\sim 1.7\times 10^{47}~{\rm erg}
\label{Epredicted}
\ee
concentrated in a size given by the corresponding Schwarzschild radius 
\be
R=\frac{2Gm}{c^2}\sim .02~ {\rm cm}
\label{R}
\ee
We may expect two main component of the signal from such an explosion: (i) a lower energy signal at a wavelength of the order of the size of the exploding object.  (ii) a higher energy signal which depend on the details of the liberated hole content.  We discuss the first signal in Section \ref{II}, the possibility of identifying it with observed signals in Section \ref{III}, and the second in Section \ref{IV}. 

\section{Low energy signal}\label{II}

A strong explosion in a small region should emit a signal with a wavelength of the order of the size of the region or somehow larger, and convert some fraction of its energy in photons.  Therefore it is reasonable to expect from this scenario an electromagnetic signal emitted in the infrared 
\be
\lambda_{predicted}\gtrsim .02~{\rm cm}.
\label{Rpr}
\ee
The received signal is going to be corrected by standard cosmological redshift. However, signals coming form farther away were originated earlier, namely younger, and therefore less massive, holes, giving a peculiar decrease of the emitted wavelength with distance. The received wavelength, taking into account both the expansion of the universe and the change of time available for the black hole to bounce, can be obtained folding \eqref{tau} into the relation between redshift and proper time.  This gives
\begin{eqnarray}
\lambda_{obs}\!&\sim&\! \frac{2Gm}{c^2} (1+z)\ \ \times \\ \nonumber
&& \  \sqrt{\frac{H_0^{-1}}{6\,k\Omega_\Lambda^{\,1/2}
}\ \sinh^{-1}\!\!\left[ \left(\frac{\Omega_\Lambda}{\Omega_M}\right)^{\!1/2} (z + 1)^{-3/2}\right]} .
\label{redhisft}
\end{eqnarray}
where we have reinserted the Newton constant $G$ and the speed of light $c$ while $H_0, \Omega_\Lambda$ and $\Omega_M$ are the Hubble constant, and the cosmological-constant and matter  densities.  Interestingly this is a very slowly varying function of the redshift. The redshift slightly over-compesates for the effect of the hole's age. The signal varies by less than an order of magnitude for redshifts up to the decoupling time (z=1100). See Figure 1.
\begin{figure}
\includegraphics[height=4cm]{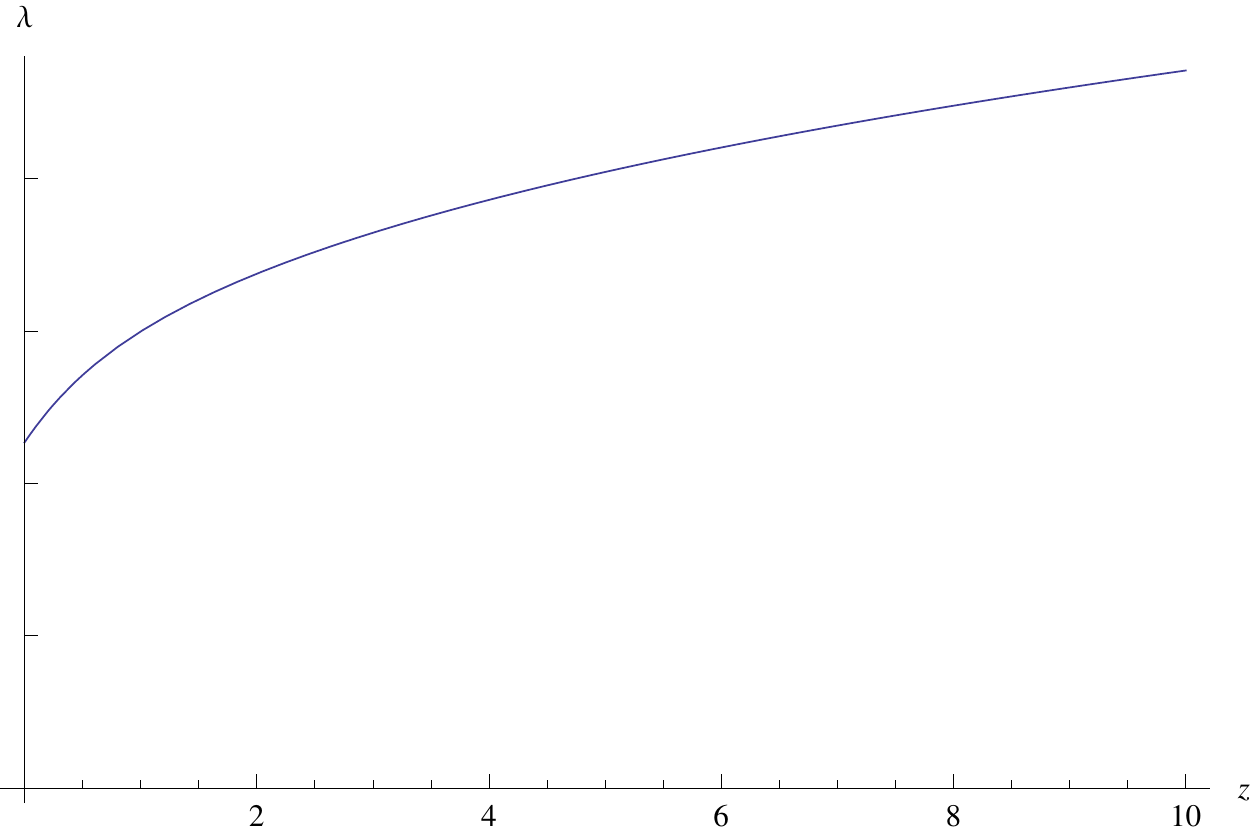}
\caption{White hole signal wavelength  (unspecified units) as a function of $z$. Notice the characteristic flattening at large distance: the youth of the hole compensate for the redshift.}
\end{figure}
If the redshift of the source can be estimated by using dispersion measures (or by identifying a host galaxy) this would be a smoking gun evidence for the phenomenon.

Do we have detectors for these signals? There are detectors operating at such wavelengths, beginning by the recently launched Herschel instrument. The 200 micron range can be observed both by PACS and SPIRE. The former employs four detector arrays, two bolometer arrays and two Ge:Ga photoconductor arrays. The latter is a camera associated with a low to medium resolution spectrometer complementing PACS. It comprises an imaging photometer and a Fourier Transform Spectrometer (FTS), both of which use bolometer detector arrays. The predicted signal falls in between PACS and SPIRE sensitivity zones. There is also a very high resolution heterodyne spectrometer, HIFI, onboard Herschel, but this is not an imaging instrument, it observes a single pixel on the sky at a time. 

However, the bolometer technology makes detecting short white-hole bursts difficult. Cosmic rays cross the detectors very often and induce glitches that are removed from the data. Were physical IR bursts due to bouncing black hole registered by the instrument, they would most probably have been flagged and deleted, mimicking a mere cosmic ray noise.  

There might be room for improvement. It is not impossible that the time structure of the bounce could lead to a characteristic time-scale of the event larger than the response time of the bolometer. In that case, a specific analysis should allow for a dedicated search of such events. We leave this study for a future work as it requires astrophysical considerations beyond this first investigation. An isotropic angular distribution of the bursts, signifying their cosmological origin, could also be considered as evidence for the model. In case many events were measured, it would be important to ensure that there is no correlation with the mean cosmic-ray flux (varying with the solar activity) at the satellite location.

Let us turn to something that \emph{has} been observed.

\section{Fast Radio Bursts}\label{III}

Fast Radio Bursts are intense isolated astrophysical radio signals with a duration of milliseconds. A small number of these were initially detected only at the Parkes radio telescope \cite{Lorimer2007,Keane2012,Thornton2013a}. Observations from the Arecibo Observatory have confirmed the detection \cite{Spitler2014}.   The frequency of these signals is in the order of $1.3$~GHz, namely a wavelength of 
\be
\lambda_{observed}\sim 20~{\rm cm}.
\label{Rob}
\ee
These signals are believed to be of extragalactic origin, mostly because the observed delay of the signal arrival time with frequency agrees quite well with the dispersion due to a ionized medium, expected from a distant source.  The  total energy emitted  in the radio by a source is estimated to be of the order $10^{38} $ erg.  The progenitors and physical nature of the Fast Radio Bursts are currently unknown \cite{Spitler2014}.  

There are three orders of magnitude between the predicted signal \eqref{Rpr} and the observed signal \eqref{Rob}.  But the black-to-white hole transition model is still very rough. It disregards rotation, dissipative phenomena, anisotropies, and other phenomena, and these could account for the discrepancy. 

In particular, astrophysical black holes rotate: one may expect the centrifugal force to lower the attraction and bring the lifetime of the hole down. In turn, this should allow larger black holes to be exploding today, and signals of larger wavelength. Furthermore, we have not taken the astrophysics of the explosion  into account.  (The total energy  \eqref{Epredicted} available in the black hole according to the theory is largely sufficient  --9 orders of magnitude larger-- than the total energy emitted in the radio estimated by the astronomers.) 

Given these uncertainties, the hypothesis that Fast Radio Burst could originate from exploding white holes is tempting, and we think deserves to be explored.

\section{High energy signal}\label{IV}

When a black hole radiates by the Hawking mechanism, its Schwarzschild radius is the only scale in the problem and the emitted radiation has a typical wavelength of this size. In the model considered here, on the other hand, the emitted energy does not come from the coupling of the event horizon with the vacuum quantum fluctuations, but rather from the time-reversal of the phenomenon that formed (and possibly, filled) the black hole. Therefore the emitted signal can be characterised by another scale: that characteristic of the matter that entered in the hole. Since the proper time of the bounce \emph{inside} the black hole is very short, there is no reason to expect this to vary much during the cosmological time.

In most simple models, primordial black holes form with a mass of the order of the Hubble Mass, $M_H\approx\frac{1}{8}t$ in Planck units,  at formation time. For black holes with masses as considered in this work, that is around $10^{23}$ kg, this corresponds to a temperature of the Universe of the order of a TeV. It is  natural to assume that a fraction of the energy of the photons emitted from the bouncing hole  be of this order of magnitude. 

The bouncing hole acts as ``redshift freezing machine" for fields inside: they are emitted back at the  energy they had when absorbed. In the meanwhile, the redshift of the surrounding universe has grown tremendously. 

Gamma-ray bursts are known at much lower energies than a TeV. Although some analysis were already carried out, no burst in the TeV range has been observed to date. Even if the rather small astrophysical background in this high-energy range is excellent from the viewpoint of detection, there is, however, a major instrumental issue: TeV detectors are ground based Cherenkov telescopes and have a very narrow field of view. The probability for a burst to occur in the appropriate direction might be small. In addition, due to the absorption by the cosmic infrared background, TeV photons cannot come from far away and the horizon is quite limited. A new generation of instruments, namely the CTA experiment, is now being designed with a huge array of telescopes that could allow to monitor many portions of the sky at the same time, opening new possibilities for this search.

The redshift dependance of this signal is different from the IR/radio one. For a hole exploding at redshift $z$, corresponding to cosmic time $t$, the signal energy is given by the temperature of the universe at  formation time, which is proportional to the inverse square root of the formation time. This time is in turn  proportional to the horizon mass which is (roughly) equal to the formation mass of the black hole. The emission wavelength is therefore proportional to the square root of the mass of the black hole. This gives an observed wavelength 
\begin{equation}
\lambda_{obs}\propto(1+z)\left(\sinh^{-1}\!\!\left[ \left(\frac{\Omega_\Lambda}{\Omega_M}\right)^{\frac12}\! (z + 1)^{-\frac32} \right]\right)^{\frac{1}{4}}.
\end{equation}
Measuring the redshift would require to associate the observed event with a host structure, which is far from being obvious, but, in principle, this dependence on $z$ provides a specific signature.

If the fraction of the total energy as gamma-rays is denoted $x$, the number of photons radiated during the bounce will be $N_{\gamma}\sim x m/E_{\gamma}$. For $x=0.2$, as a reasonable example, this leads to $10^{46}$ $\gamma$-rays in the TeV range. If one considers an effective telescope area given by a disc of radius 100 meters (the approximate size of the Cherenkov shower) and requires 10 measured photons for each burst, the bouncing object can be detected up to a distance of $D\approx 10^{24}$ m, which is around 100 million light-years or a redshift of z=0.01. This is within the $\gamma$-ray horizon and the latter is therefore not the limiting factor. A promising strategy could be to point the telescope toward a galaxy with $z<0.01$.  If it is not a blazar, the TeV background is expected to be  small or vanishing. If bouncing primordial black holes around $10^{23}$ kg are to represent a large fraction of the dark matter, there could be as much as $10^{19}$ objects of this type within the galaxy. Each exploding (bouncing) one would be detected. Of course, the actual number of events per unit of time depends of the width of the primordial mass spectrum (if any), which is not known. But orders of magnitude show that detection is not hopeless.

\section{Conclusion}

We have discussed the signal of a primordial black hole exploding today via a black-to-white quantum transition \cite{Haggard2014} and the possibility of observing the lower as well as the higher energies components of the signal.  We have observed that the first would have a characteristic distance-frequency relation flattening at large redshift.  

We have pointed out the possibility of identifying this signal with the Fast Radio Bursts observed by the Arecibo and Parkes observatories. 

A  connection between black hole explosions and short radio signals was suggested time ago by M. Rees  \cite{Rees1977}.  The physics considered by Rees is different from that considered here: radio or optical emission from the relativistic shock wave generated from the explosion of \emph{small} black holes, interacting with an ambient magnetic field. 

In the scenario we have considered here, on the contrary, the phenomenon is of direct quantum gravitational nature. A quantum gravitational phenomenon can have effects at observable scales because the presence of the large multiplicative factor  \cite{Amelino-Camelia2013}
\be
\frac{t_H}{t_P}\sim 8\times 10^{60}.
\ee
in the physics of the phenomenon. 

If the observed Fast Radio Bursts are connected to this phenomenon, they represent the first known direct observation of a quantum gravitaty effect.

\vskip.5cm 

We thank Massimo Cerdonio for suggestions and advices. FV acknowledges support from the Netherlands Organisation for Scientific Research (NWO) Veni Fellowship Program, and from the Centre National de la Recherche Scientifique (CNRS) visiting program for supporting her visit at the CPT in Marseille.

\end{document}